\documentclass[prd, amsfonts, twocolumn, nofootinbib, showpacs]{revtex4}
\usepackage{graphicx, epsfig}
\usepackage{color}
\usepackage{amsmath}
\usepackage{amssymb}
\newcommand{\be}{\begin{equation}}
\newcommand{\ee}{\end{equation}}
\newcommand{\bea}{\begin{eqnarray}}
\newcommand{\eea}{\end{eqnarray}}

\newcommand{\gapp}{\mathrel{\raise.3ex\hbox{$>$}\mkern-14mu
\lower0.6ex\hbox{$\sim$}}}
\newcommand{\lapp}{\mathrel{\raise.3ex\hbox{$<$}\mkern-14mu
\lower0.6ex\hbox{$\sim$}}}
\def\bbox{{\,\lower0.9pt\vbox{\hrule \hbox{\vrule height 0.2 cm
\hskip 0.2 cm \vrule  height 0.2 cm}\hrule}\,}}

\begin{document}
\title{A Vector type of Unruh-DeWitt-like detector}
\author{De-Chang Dai}
\affiliation{ Institute of Natural Sciences, Shanghai Key Lab for Particle Physics and Cosmology, and Center for Astrophysics and Astronomy, Department of Physics and Astronomy, Shanghai Jiao Tong University, Shanghai 200240, China}


\begin{abstract}
\widetext
We study a type of an Unruh-DeWitt-like detector based on a vector rather than scalar field. This detector has two energy states and produces Larmor radiation when there is no energy gap between them. This setup indicates that Larmor radiation and Unruh radiation are two counterparts of the same phenomenon. Larmor radiation is observed  in the inertial frame, while Unruh radiation is observed in an accelerated frame. The accelerated observer sees that his detector absorbed a particle inside its own accelerated horizon, while the Larmor radiation is the companion particle which leaves the accelerated observer's horizon. Since the detection is based on the electromagnetic field, this type of detector is much closer to the real world than a standard Unruh-DeWitt detector which is coupled to a scalar field.
\end{abstract}


\maketitle

\section{introduction}
Unruh-Dewitt detector is a theoretical setup initially introduced by Unruh and DeWitt\cite{Unruh:1976db,DeWitt,Unruh:1983ms,Higuchi:1992td,Higuchi:1992we,Higuchi:1993fn} to study phenomena of quantum fields in curved spacetimes, or equivalently phenomena  observed by an accelerated observer. Among the other things, this setup reveals that a uniformly accelerated observer in Minkowski space should observe a thermal spectrum\cite{Unruh:1976db} with temperature $T_u=\frac{a}{2\pi}$, where $a$ is the acceleration. This thermal radiation is called the Unruh radiation. Equivalently,  one can use the same setup to study thermal radiation seen by a static detector in Schwarzschild \cite{Hartle:1976tp} and de-Sitter \cite{Gibbons:1977mu} spacetimes.

The root of Unruh radiation is acceleration. Since an eternally accelerating observer observes  a horizon (a region of spacetime from which no signal can emerge), Unruh radiation is considered to be equivalent to Hawking radiation which is caused by a black hole event horizon. Since it is very difficult (if possible at all) to create a black hole in a laboratory to observe Hawking radiation, Unruh radiation can serve as a good alternative to study aspects of Hawking radiation. Recently, it was proposed that a high intensity laser can test the Unruh radiation through its strong electric field which causes acceleration of a large magnitude \cite{Chen:1998kp,Thirolf,Schutzhold:2006gj}. However, this proposal is not straightforward, and different studies disagree about the existence of Unruh radiation created in this way \cite{1991RSPSA.435..205R,Iso:2010yq,Raval:1995mb,Iso:2013sm,Oshita:2015qka}. More work is certainly needed to clarify all the issues.

One of the problems in the original proposal is that the model in question was based on a scalar field. In contrast, when studying radiation from electrons one has to consider photons, which are vector particles. However, so far only two types of Unruh-DeWitt-like detectors were constructed. One is a detector coupled to a scalar field \cite{Unruh:1976db,DeWitt}, while the other is coupled to the derivative of a scalar field\cite{1991RSPSA.435..205R}. In this paper we extend the construct to a vector field. In our analysis, we reproduce the Larmor radiation in the non-relativistic regime. This proves that Larmor radiation is basically Unruh radiation. They are the same phenomena observed by different observers. This result agrees with earlier studies \cite{Pena:2014uia}. We then apply the detector setup to a long time uniformly accelerated charged detector. Our result shows that the spectrum is related to Unruh temperature $T=\frac{a}{2\pi}$, but its form is different from the scalar field. 


\section{Unruh DeWitt detector}

Unruh De-Witt detector is considered to be a point-like detector which moves in a relativistic spacetime on a smooth path $\mathbf{x}(\tau)$ (where $\tau $ is the detector's proper time). This detector usually consists of the ground state ($|0>_D$) and an excited state $|1>_D$, with eigen energies $0$ and $\omega$ respectively. The detector is usually coupled to a scaler field, $\phi$, or the derivative of the scalar field, $\partial_\tau \phi$. As an example, here we couple it  to $\phi$ for simplicity. The interaction is described by the following Hamiltonian
\begin{equation}
H_{int}=c\mu (\tau )\phi (\mathbf{x}(\tau)) .
\end{equation}
Here, $c$ is a small coupling constant and $\mu$ is a monopole moment operator.  The probability of the transition rate from $|0>_D$ to $|1>_D$  can be found from the first order perturbation theory
\begin{equation}
P(\omega) =c^2 |_D<1|\mu(0)|0>_D|^2 F(\omega).
\end{equation}
Here, $F(\omega)$ is the detector's response function
\begin{eqnarray}
F(\omega)=\int^\infty_{-\infty} d\tau d\tau' e^{-i\omega (\tau-\tau')}<0|\phi(\mathbf{x}(\tau))\phi(\mathbf{x}(\tau'))|0>
 \end{eqnarray}
 $|0>$ is $\phi$'s ground state. The unit time response is
 \begin{eqnarray}
\dot{F}(\omega)|_\tau=\int^\infty_{-\infty} d\tau' e^{-i\omega (\tau-\tau')}<0|\phi(\mathbf{x}(\tau))\phi(\mathbf{x}(\tau'))|0>
 \end{eqnarray}

If the detector undergoes a uniform linear acceleration $a$ in $z$-direction as
\begin{eqnarray}
\label{x1}
t&=&\frac{\sinh(a\tau)}{a}\\
\label{x2}
z&=&\frac{\cosh(a\tau)}{a} \\
\label{x3}
x&=&y=0
 \end{eqnarray}
, then the unit time response becomes
 \begin{equation}
 \dot{F}(\omega)=\frac{1}{2\pi}\frac{\omega}{e^{2\pi\omega/a}-1} .
 \end{equation}
From here we can read-off the temperature $T=\frac{a}{2\pi}$. Therefore, an accelerated observer sees a thermal bath radiation.

\section{ Vector type Unruh-Dewitt detector}

In this section we want to extend the construct to a vector field type of detector. This detector is coupled to a vector field, $A^\mu$, in particular the electromagnetic field. Such detector cannot be of monopole type. It must include a vector current, $j^\mu(\tau)$,  which couples to  the vector field through the following Hamiltonian
\begin{equation}
H_{int} = e A_\mu (\mathbf (\tau))j^\mu(\mathbf (\tau))
\end{equation}
Here,
\begin{eqnarray}
j^t &=&v^t(\tau)\delta (x)\delta (y) \delta (z-z(\tau))\\
j^z &=&v^z(\tau)\delta (x)\delta (y) \delta (z-z(\tau))\\
j^x&=&j^y=0
\end{eqnarray}
where $v^\alpha$ is the four velocity. We consider the detector which moves in z-direction only, and has two eigenstates, $|0>_j$ and $|1>_j$, with eigen energies $0$ and $\omega$ respectively. The probability of transition rate from $|0>_j$ to $|1>_j$  can again be found from first order perturbation theory

\begin{eqnarray}
P_j(\omega)&=&e^2\int^\infty_{-\infty} d\tau d\tau' v_\alpha(\tau)v_\beta(\tau')e^{-i\omega (\tau-\tau')}\nonumber\\
\label{energy}
&\times&<0\gamma|A^\alpha(\mathbf{x}(\tau))A^\beta(\mathbf{x}(\tau'))|0\gamma>,
 \end{eqnarray}
where  $|0\gamma>$ is $A^\mu$'s ground state.

\begin{eqnarray}
&&<0\gamma|A^\alpha(\mathbf{x}(\tau))A^\beta(\mathbf{x}(\tau'))|0\gamma>\nonumber\\
&&= \int \frac{d^3\vec{k}}{2k (2\pi)^3}\epsilon^\alpha(k) \epsilon ^\beta (k)e^{-i\mathbf{k}\cdot (\mathbf{x}(\tau))-\mathbf{x}(\tau'))} .
\end{eqnarray}
Here, $\epsilon^\alpha(k)$ is the polarization of the vector field, and it is transverse to $k^\mu$. The term $\epsilon^\alpha(k) \epsilon ^\beta (k)$ should not be simplified to $\eta^{\alpha\beta}$ since it is the external field which includes only two degree of freedoms.

\section{Larmor radiation}

\begin{figure}[ht!]
   \centering
\includegraphics[width=6cm]{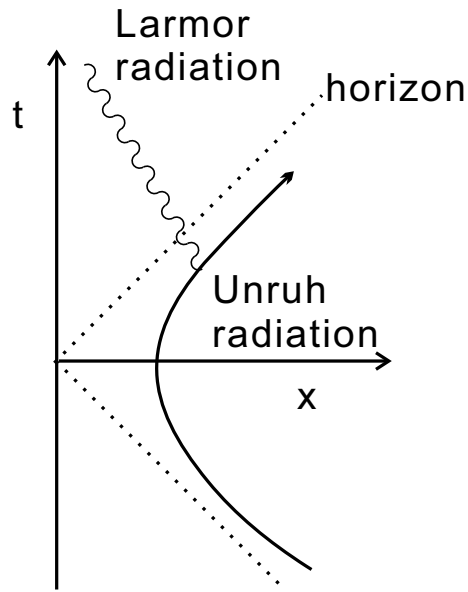}
\caption{An accelerated particle creates a horizon in its comoving frame. This particle emits a vector field which passes through its horizon and becomes Larmor radiation. However, the comoving observer treats the same vector field as the Unruh radiation.}
\label{radiation}
\end{figure}

We now consider a small magnitude acceleration in a non-relativistic limit, i.e. $\Delta x<<\Delta t$ and $\Delta t\approx \Delta\tau$. The surviving term in this limit in equation \ref{energy} is the term which includes only the z-component
\begin{eqnarray}
&&<0\gamma|A^z(\mathbf{x}(\tau))A^z(\mathbf{x}(\tau'))|0\gamma>\nonumber\\
&&\approx \frac{1}{6\pi^2} \int_0^\infty  ke^{-ik(\tau-\tau')}dk
\end{eqnarray}
If $\omega =0$, then equation \ref{energy} becomes
\begin{eqnarray}
&&P_j=\frac{e^2}{6\pi^2} \int_0^\infty \frac{|\tilde{a}(k)|^2}{k} dk\\
&&\tilde{a}(k)=\int a(t) e^{-ik t}dt
\end{eqnarray}
This is nothing else but non-relativistic Larmor radiation \cite{Higuchi:2002qc}. This implies that Larmor radiation is just a special case of Unruh effect. The difference is that Larmor radiation is observed in Minkowski spacetime, while Unruh effect is observed in a co-moving spacetime( fig. \ref{radiation}). Indeed, we can arrive to the same conclusion from the equivalence principle. Larmor radiation and Unruh radiation are just two complementary parts of the same phenomenon - Unruh radiation is observed by the accelerated observer inside his own horizon, while Larmor radiation is the radiation that leaves the observer's horizon. Larmor radiation is then detected by a non-accelerated observer in Minkowski space.

\section{Long time uniform acceleration}
We now consider the detector which is uniformly accelerated for long time but not eternally. Its trajectory can be approximately described by equations \ref{x1} to \ref{x3}. In this case
\begin{eqnarray}
&&<0\gamma|A^z(\mathbf{x}(\tau))A^z(\mathbf{x}(\tau'))|0\gamma>\nonumber\\
&&\approx \frac{-1}{6\pi^2} \int_{-\infty}^\infty  \frac{ke^{-ik(\tau-\tau')}}{e^{2\pi k/a}-1}dk
\end{eqnarray}
The denominator is in the Planckian form, so $T=\frac{a}{2\pi}$ can be treated as a thermal temperature. The transmission rate, equation \ref{energy}, becomes
\begin{eqnarray}
\label{radiation2}
&&P_j=-\frac{e^2}{6\pi^2} \int_{-\infty}^\infty \frac{|\tilde{a}(k+\omega)|^2}{(k+\omega)^2}\frac{k}{e^{2\pi k/a}-1} dk
\end{eqnarray}
The response depends also on  $\omega+k$. This implies that the energy, $k+\omega$, is extracted from the accelerating source. $\omega$ part is absorbed by the detector and $k$ is released to the event horizon and disappears from detector's side. It becomes Larmor radiation after it cross the horizon. One finds that equation \ref{radiation2} reduces to non-relativistic Larmor radiation as $a\rightarrow 0$. However, the Unruh temperature is related to $k$ instead of $\omega$, which is what a scalar type detector predicts. Similar result has been found in two energy-level-atom case\cite{Zhu:2006wt}.

\section{Conclusion}
We reviewed the standard Unruh-DeWitt detector and extend it to a vector type of Unruh-DeWitt-like detector. This new detector can give more realistic results in the context of radiation from a charged accelerated device. If the detector does not have an energy gap between its eigen states, the result is the same as the Larmor radiation. This confirms that Unruh radiation and Larmor radiation are two parts of the same phenomenon which are observed by different observers\cite{Pena:2014uia}. Unruh radiation is seen by an accelerated observer and Larmor radiation is seen by an inertial frame observer. In other words, Unruh radiation is observed within the accelerated observer's horizon, while Larmor radiation is the part that falls into the accelerated observer's horizon.

We also considered  Unruh radiation for a long-time (but finite) uniformly accelerated observer. The effect is shown in equation \ref{radiation2}. Radiation is still characterized by Unruh temperature $T=\frac{a}{2\pi}$, but also depends on $\omega +k$. This implies that some of the energy is extracted from the accelerated source to the excited detector. This energy is radiated outside of the horizon (Larmor radiation part).

\begin{acknowledgments}
D.C. Dai was supported by the National Science Foundation of China (Grant No. 11433001 and 11447601), National Basic Research Program of China (973 Program 2015CB857001), No.14ZR1423200 from the Office of Science and Technology in Shanghai Municipal Government, the key laboratory grant from the Office of Science and Technology in Shanghai Municipal Government (No. 11DZ2260700) and  the Program of Shanghai Academic/Technology Research Leader under Grant No. 16XD1401600.

\end{acknowledgments}

\end{document}